\documentclass[preprint,showpacs,preprintnumbers,amsmath,amssymb]{revtex4}

\usepackage{graphicx}
\usepackage{dcolumn}
\usepackage{bm}

\begin{document}

\preprint{APS/123-QED}

\title{On the elliptical flow and asymmetry of the colliding nuclei   
\\}

\author{Varinderjit Kaur}
\author{Suneel Kumar}%
\email{suneel.kumar@thapar.edu}

\affiliation{%
School of Physics and Materials Science, Thapar University Patiala-147004, Punjab (India)\\
}
\author{Rajeev K. Puri}%
\affiliation{
Department of Physics, Panjab University, Chandigarh (India)\\
}%

\date{\today}
\begin{abstract}
A study of elliptical flow is presented with respect to the asymmetry of colliding nuclei using the 
reactions of $_{24}Cr^{50}+_{44}Ru^{102}$, $_{16}S^{32}+_{50}Sn^{120}$ and $_{8}O^{16}+_{54}Xe^{136}$ 
at incident energies between 50 and 250 MeV/nucleon within the framework of isospin-dependent quantum 
molecular dynamics model. For the present analysis, total mass of the colliding pairs is kept fixed and 
asymmetry is varied as ${\eta}$ = 0.2, 0.3, 0.7. The elliptical flow shows a transition 
from in-plane to out-of-plane in the mid rapidity region with incident energy. The transition energy is 
found to increase with the asymmetry for lighter fragments.
\end{abstract}
\maketitle
\baselineskip=1.0\baselineskip
\section{Introduction}
Large efforts are going on to understand the nuclear matter at the extreme conditions of temperature
and density and also to 
explore the role of symmetry energy under these extreme conditions \cite{report}. One of the most 
sought after phenomena in this direction is the collective flow and its various forms 
\cite{westfall,tsang,sood2006,chen}. For the last few years, collective flow has been used as a powerful 
useful tool to explore the nuclear equation of state (EOS) as well as in-medium nucleon-nucleon 
cross-sections \cite{report,reisdorf}. Collective flow is a motion characterized
by the space-momentum correlations of dynamic origin. Following two different signatures of collective 
flow have been predicted: a) bounce-off of compressed matter in the reaction plane \cite{v1} and
b) squeeze-out of the participant matter out of the reaction plane \cite{v2}.\\
Such observables together represent the anisotropic part of the transverse flow that appears
in the non-central heavy-ion collisions only. The highly stopped and compressed nuclear matter around the 
mid-rapidity region is seen directly in the squeeze out \cite{hartnack}, also known as elliptical flow. 
The elliptical flow has been proven to be one of the most fruitful probes to study the dynamics 
of heavy-ion collisions. The elliptical flow describes the eccentricity  of an ellipse-like distribution. 
Quantitatively, it is the difference between the minor and major axis. The orientation of the major 
axis is confined to the azimuthal angle ${\phi}$ or ${\phi+\frac{\pi}{2}}$ for an ellipse-like 
distribution. The major axis lies within the reaction plane for ${\phi}$; whereas ${\phi+\frac{\pi}{2}}$  
indicates that the orientation of the ellipse is perpendicular to the reaction plane (i.e., 
squeeze-out flow). The parameter of the elliptical flow is quantified by the second order Fourier 
coefficient ${<v_2>} = ({\Large <}\frac{p_x^{2}-p_y^{2}}{p_x^{2}+p_y^{2}}{\Large >}$, ${p_x}$ and 
${p_y}$ being the x and y components of the momentum respectively), from the 
azimuthal distribution of detected particles at mid-rapidity \cite{voloshin}:\\
$$\frac{dN}{d\phi} = p_0( 1+2v_{1}Cos{\phi}+2v_{2}Cos2{\phi}).$$\\
where ${\phi}$ is the azimuthal angle between the transverse momenta of the particles and reaction plane. 
The positive values of elliptical flow reflect an in-plane emission, whereas out-off plane emission is
reflected by the negative values. The parameters ${<Cos2\phi>}$ of elliptical flow depend on the complex 
interplay between the expansion, rotation and shadowing of the spectators, apart from the incident 
energy. It is worth mentioning that both the mean field and binary nucleon-nucleon collision parts play an
important role at intermediate energies. The mean field potential plays a dominant role at low 
incident energies, which is gradually taken over by the two body collision part at higher incident
energies. Therefore, a detailed study on the excitation function of elliptical flow in this energy 
range can provide useful information about the nucleon-nucleon interactions and origin of the isospin 
effects in heavy-ion collisions.\\ 
It is worth mentioning that the outcome of a reaction depends also on the asymmetry of the reaction. 
Unfortunately, very few investigations focus on this aspect. The asymmetry of a reaction can be defined 
by the asymmetry parameter ${{\eta}=  {(A_T-A_P)}/{(A_T+A_P)}}$; \cite{asymm} 
where ${A_T}$ and ${A_P}$ are the masses of the
target and projectile, respectively. The ${\eta}$ = 0 corresponds to the symmetric reactions, 
whereas non-zero values of ${\eta}$ define different asymmetries of a reaction. As noted by FOPI 
group \cite{diogene,plastic,nautilus,eos,LAND,sanjeev014611}, the 
reaction dynamics in a symmetric reaction (${\eta}$ = 0) can be quite different compared to an 
asymmetric reaction (${{\eta} \ne 0}$). This is valid both at low and intermediate energies.
This difference emerges due to the different deposition of the excitation energy (`in form of 
compressional 
and thermal energies) in symmetric and asymmetric reactions. Though the systematic role of asymmetry 
has been explored in multifragmentation, no such study yet exists in the literature for elliptical flow 
\cite{PRC58}. We plan to address this in present paper.\\
We plan to understand how the elliptical flow is affected by the asymmetry of a reaction. 
This study is performed within the
isospin-dependent quantum molecular dynamics (IQMD) model discussed in section II. Our
results are presented in section III. Finally, we summarize the results in section IV.
\section{The Model}

The isospin-dependent quantum molecular dynamics (IQMD)\cite{sanjeev014611,PRC58,hartnack1} model treats 
different charge states of nucleons, deltas and pions explicitly, as inherited from the Vlasov-Uehling-
Uhlenbeck (VUU) model. The IQMD model has been used successfully for the analysis of a large number of 
observables from low to relativistic energies. The isospin degree of freedom enters into the calculations 
via symmetry potential, cross sections and Coulomb interactions.  \\
In this model, baryons are represented by Gaussian-shaped density distributions \\
\begin{equation}
f_i(r,p,t) = \frac{1}{{\pi}^2{\hbar}^2}e^{\frac{{-(r-r_i(t))^2}}{2L}}e^{\frac{{-(p-p_i(t))^2}.2L}{\hbar^2}}.
\end{equation}
where L is the Gaussian width which is taken to be 2.16 ${fm^2}$. In ref \cite{hartnack1}, this Gaussian width is found to be dependent
on the size of the system. 
Nucleons are initialized in a sphere with radius R = ${1.12A^{1/3}}$ fm, in accordance with the  liquid 
drop model. Each nucleon occupies a volume of ${\hbar^3}$ so that phase space is uniformly filled.
The initial momenta are randomly chosen between 0 and Fermi momentum ${p_F}$. The nucleons of the target 
and projectile interact via two and three-body Skyrme forces, Yukawa potential and Coulomb interactions.
The isospin degrees of 
freedom is treated explicitly by employing a symmetry potential and explicit Coulomb forces between 
protons of the colliding target and projectile. This helps in achieving the correct
distribution of protons and neutrons within the nucleus.\\
The hadrons propagate using Hamilton equations of motion:\\
\begin{equation}
\frac{d\vec{r_i}}{dt} = \frac{d<H>}{d{p_i}}~~~~;~~~~\frac{d\vec{p_i}}{dt} = -\frac{d<H>}{d{r_i}}.
\end{equation}
 with\\ 
${ <H> = <T> + <V>}$  is the Hamiltonian, which is written as:
\begin{eqnarray}
    =  \sum_i\frac{p_i^2}{2m_i} + \sum_i \sum_{j>i}\int f_i(\vec{r},\vec{p},t)V^{ij}(\vec{r'},\vec{r})\nonumber\\ 
f_j(\vec{r'},\vec{p'},t)d\vec{r}d\vec{r'}d\vec{p}d\vec{p'}.
\end{eqnarray}
The baryon-baryon potential ${V^{ij}}$, in the above relation, reads as\\
\begin{eqnarray}
V^{ij}(\vec{r'}-\vec{r}) &~=~& V_{Skyrme}^{ij} + V_{Yukawa}^{ij} + V_{Coul}^{ij} + V_{Sym}^{ij}\nonumber\\
&=&t_1\delta(\vec{r'}-\vec{r})+ t_2\delta(\vec{r'}-\vec{r}){\rho}^{\gamma-1}(\frac{\vec{r'}+\vec{r}}{2})\nonumber\\ 
&+& t_3\frac{exp({\mid{\vec{r'}-\vec{r}}\mid}/{\mu})}{({\mid{\vec{r'}-\vec{r}}\mid}/{\mu})} + \frac{Z_{i}Z_{j}{e^2}}{\mid{\vec{r'}-\vec{r}\mid}}\nonumber\\
&+& t_{4} \frac{1}{\rho_o}T_{3}^{i}T_{3}^{j}.\delta(\vec{r'_i} - \vec{r_j}).
\end{eqnarray}
Where ${{\mu} = 0.4 fm}$, ${t_3 = -6.66 MeV}$ and ${t_4 = 100 MeV}$. The values of ${t_1}$ and ${t_2}$ 
depends on the values of ${\alpha}$, ${\beta}$, and ${\gamma}$ \cite{report}.
Here ${Z_i}$ and ${Z_j}$ denote the charges of the ${i^{th}}$ and ${j^{th}}$ baryon, and ${T_{3}^i}$, 
${T_{3}^j}$ are their respective ${T_3}$ components (i.e. 1/2 for protons and -1/2 for neutrons). 
The Meson potential consists of Coulomb interaction only. The parameters ${\mu}$ and ${t_1,........,t_4}$ 
are adjusted to the real part of the nucleonic optical potential. \\
\section{Results and Discussions}
For a controlled study of the role of asymmetry of a reaction, the total reacting mass is fixed equal to 
152 units. While total mass stayed constant, asymmetry ${\eta}$ is varied by chosing different 
combinations of projectile-target. We shall perform exclusive studies by simulating the reactions of  
$_{24}Cr^{50}+_{44}Ru^{102}$ (${\eta = 0.3}$), $_{16}S^{32}+_{50}Sn^{120}$ (${\eta = 0.5}$), 
and $_{8}O^{16}+_{54}Xe^{136}$ (${\eta = 0.7}$) at incident energies between 50 and 
250 MeV/nucleon for semi-central impact parameter using a soft equation of state. The phase space 
generated by the IQMD model is analyzed using the minimum spanning tree (MST) \cite{report} method. 
This method binds two nucleons in a fragment if their 
distance is less than 4 fm. \\
As stated above, positive value of elliptical flow describes the eccentricity of an 
ellipse-like distribution and indicates in-plane enhancement of the particle emission, i.e., 
rotational behavior. On the other hand, a negative value of ${<v_2>}$ shows the squeeze-out effects
perpendicular to the reaction plane. Obviously, zero value corresponds to an isotropic distribution.
Generally, for a meaningful understanding ${v_2}$ is extracted from the midrapidity region only.
Naturally, midrapidity region corresponds to the collision (participant) 
zone and hence signifies compressed matter. On the other hand, ${Y_{c.m.}/Y_{beam}}$ ${\ne}$ 0  
corresponds to spectator region; ${Y_{c.m.}/Y_{beam} < -0.1}$ corresponds to target like (TL) 
matter whereas ${Y_{c.m.}/Y_{beam} > 0.1}$ corresponds to projectile like (PL) matter.\\
In Fig.1, final state elliptical flow is displayed for free particles (upper panel) and    
light charged particles (LCP's) [${(2\le A \le4)}$] (lower panel) as a function of transverse momentum 
${(P_t)}$. One can see a Gaussian shaped behavior at all asymmetries quite similar to the one reported 
by Colona and Toro {\it et al.}, \cite{colona}. Note that the Gaussian shaped behavior is integrated over the 
entire rapidity range. One also sees that elliptical flow of nucleons/LCP's is 
positive over entire ${P_t}$ range. One also notices that for larger asymmetries (e.g. ${\eta = 0.7}$), 
prominent peaks of Gaussian noted earlier diminishes. There is a linear increase in the elliptical flow with transverse 
momentum upto certain incident energy. Obviously, particles with larger momentum will escape the reaction 
zone at an earlier time. 
After certain value, elliptical flow decreases indicating that not enough particles occupy such higher 
${P_t}$ values. One also notices lower peaks of Gaussian for heavier fragments. The reason for this 
shift is that the emitted free nucleons can feel 
the role of mean-field directly, while the LCP's have a weaker sensitivity \cite{yan}. 
Interestingly, ${<v_2>}$ depends significantly on the symmetry of the reaction also.  
\begin{figure}
\includegraphics{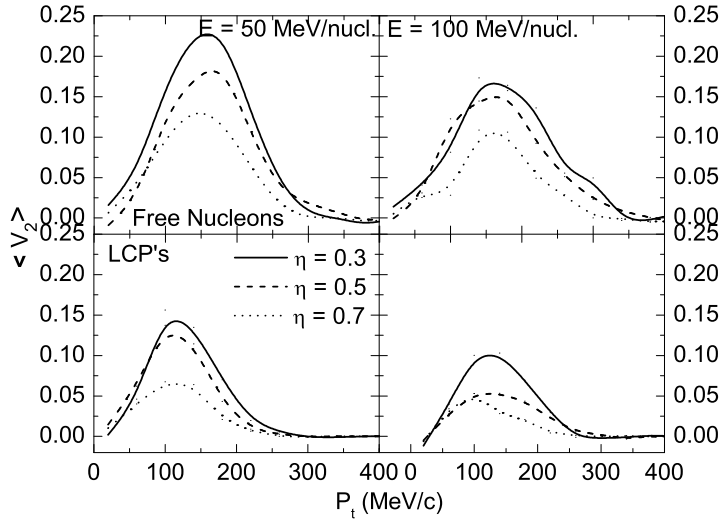}
\caption{\label{fig:1} The transverse momentum dependence of elliptical flow, summed over 
all rapidity bins at $\hat{b}$ = 0.3 for different asymmetries at 50 (left) and 100 (right)
 MeV/nucleon. The upper and lower panels represent the free nucleons and light charged particles (LCP's),
respectively.}  
\end{figure}
Further, we see that neutron-rich system ($_{8}O^{16}+_{54}Xe^{136}$ with N/Z = 1.4) exhibits a 
weaker squeeze-out flow compared to other reactions. These findings are in agreement with the one 
reported by Zhang {\it et al.}, \cite{zhang}. Moreover, the N/Z effect is more pronounced at 
E = 50 MeV/nucleon. This also indicates that in addition to the mass asymmetry, the mean field effects such as
isospin effects are also responsible for different elliptical flow.   \\
In Fig. 2, we divide the total elliptical flow into contributions from target-like (TL), mid-rapidity, 
and projectile-like (PL) particles. From the figure, we see that the projectile-like (PL) nucleons
and LCP's feel more squeeze out compared to target-like (TL) nucleons/LCP's. Due to larger 
asymmetry, only small fraction of nucleons/LCP's experience squeeze out compared to symmetric 
reactions. This decrease of squeeze out with asymmetry happens due to decreasing participant zone.
This is in agreement with earlier calculations where fragments were found to exhibit similar trends.\\  

\begin{figure}
\includegraphics{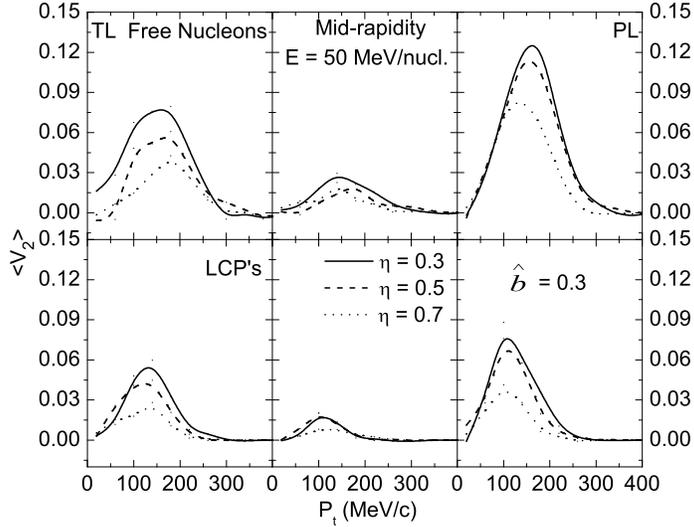}
\caption{\label{fig:2} The transverse momentum dependence of the elliptical flow at E =
50 MeV/nucleon for different asymmetries divided into contributions from target-like, 
midrapidity and projectile-like matter, respectively; the upper and lower panels have 
same meaning as in Fig. 1.}  
\end{figure}
Since the change of the asymmetry of a reaction can affect the elliptical flow via symmetry energy, 
we display in Fig. 3, the transverse momentum dependence of elliptical flow for LCP's in the 
mid-rapidity region for the reaction of $_{24}Cr^{50}+_{44}Ru^{102}$ with and without symmetry energy 
The effect of symmetry 
energy is clearly visible. This is in agreement with the findings of Chen {\it et al.},
\cite{chen1}, where it was concluded that the production of LCP's act as a probe for
symmetry energy. \\
\begin{figure}
\includegraphics{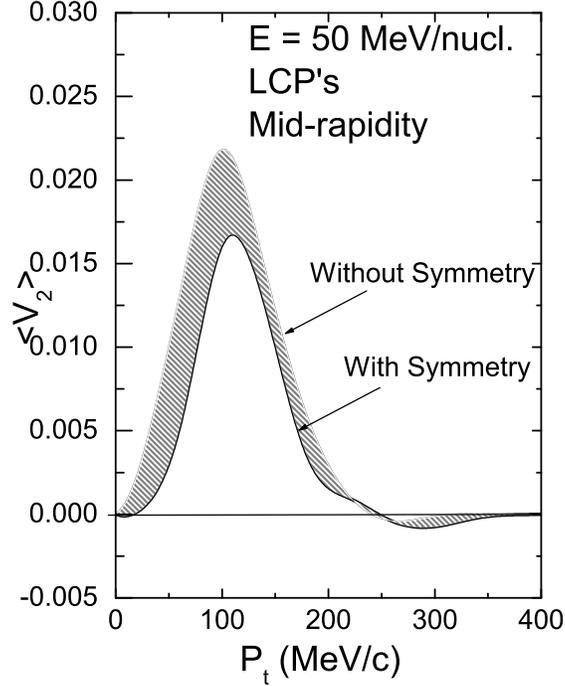}
\caption{\label{fig:3} The transverse momentum dependence of elliptical flow for LCP's in 
the midrapidity region at E = 50 MeV/nucleon. The panel exhibits the effect of symmetry energy on 
$_{24}Cr^{50}+_{44}Ru^{102}$ reaction at $\hat{b}$ = 0.3.}  
\end{figure}
In Fig. 4, we display the variation of excitation function ${<v_2>}$ for  LCP's 
as a function of incident energy for entire rapidity region and for mid rapidity region 
${-0.1 \le Y^{red} \le 0.1}$ only. The general behavior of excitation functions for various asymmetries
is quite similar. The microscopic behavior, however, depends on the asymmetry of the reaction.
Interestingly, no transition in the elliptical flow occurs when entire rapidity region is considered.
In contrast, a transition from the preferential in-plane (rotational like) emission (${{v_2} > 0}$) to 
out-of-plane (squeeze-out) emission (${{v_2} < 0}$) occurs at mid rapidity zone. This happens due
to the fact that the contribution of spectator matter increases with rapidity region, leading to less 
squeeze-out of the particles in entire rapidity region. On the other hand, the contribution of the
participant zone dominates the reaction in midrapidity region leading to the transition from in-plane 
to out-of-plane. This happens because the mean field which contributes to the
formation of a rotating compound system becomes less important and collective expansion
process (based on the nucleon-nucleon binary scattering) starts to be predominant \cite{zhang1}. The 
competition between the mean-field and nucleon-nucleon collisions should strongly depend on the 
effective interactions, which leads to the divergence of the transition energies calculated by varying the 
asymmetry of a reaction. In other words, participant zone is primarily responsible for the
transition from in-plane to out-of-plane. The energy at which this 
transition is observed is dubbed as the transition energy ${E_{trans}}$. That is why, LCP's, 
which originate from the participant zone, show a clear and systematic transition with the beam energy 
as well as with the asymmetry of a reaction.  
\begin{figure}
\includegraphics{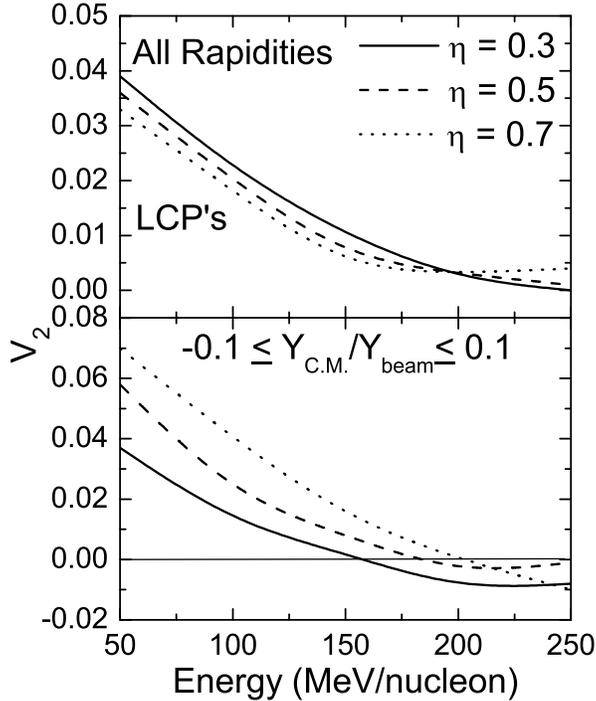}
\caption{\label{fig:4} The variation of elliptical flow (summed over the entire transverse 
momentum) with beam energy at $\hat{b}$ = 0.3 for different asymmetries over the entire rapidity range
(left panels) and at midrapidity (right panels). The upper and lower panels have same meaning as in Fig. 1.}  
\end{figure}
One should note that transition energies increases with the asymmetry of a reaction.\\
\section {Conclusion}
In the present study, elliptical flow is studied for different asymmetries leading to same compound 
masses. For this, the reactions of $_{24}Cr^{50}+_{44}Ru^{102}$ 
(${\eta = 0.3}$), $_{16}S^{32}+_{50}Sn^{120}$ (${\eta = 0.5}$), and $_{8}O^{16}+_{54}Xe^{136}$ 
(${\eta = 0.7}$) are simulated at incident energies between 50 and 250 MeV/nucleon using isospin-dependent 
quantum molecular dynamics model. The characteristic features of the elliptical flow are described by 
varying the mass asymmetry. The elliptical flow is found to show a transition 
from in-plane to out-of-plane in the mid rapidity region with incident energy. The transition energy at 
which the 
elliptical flow ${v_2}$ changes sign from positive to negative are different for different asymmetries. 
The transition energy is found to increase with the asymmetry for lighter fragments.

\section {Acknowledgment}
This work has been supported by the grant from Department of Science and Technology (DST), Government of 
India, vide Grant No.SR/WOS-A/PS-10/2008.\\ 

\end{document}